**THE EUROPEAN**
**PHYSICAL JOURNAL C**

# Obstacles from interstellar matters and distortion in warp drive superluminal travel scenario

Siyu Bian[1,a], Yi Wang[1,2,b], Zun Wang[1,c], Mian Zhu[1,2,d]

[1] Department of Physics, The Hong Kong University of Science and Technology, Clear Water Bay, Kowloon, Hong Kong, People's Republic of China
[2] Jockey Club Institute for Advanced Study, The Hong Kong University of Science and Technology, Clear Water Bay, Kowloon, Hong Kong, People's Republic of China

**Abstract**  We investigate obstacles of superluminal "warp drive" travels from interactions with interstellar matter and from curvature effects. The effect of collision of interstellar dust particles and photons with the spacecraft will all lead to a pressure proportional to the apparent velocity of the spaceship $v_s$. The force exerted on the spacecraft from the curvature effect has two non-trivial components. The radial and longitudinal components scales as $v_s^2$ and $v_s^4$ respectively. The above obstacles become increasingly important when the spaceship travels at high superluminal speeds.

## Contents

a e-mail: sbianaa@connect.ust.hk
b e-mail: phyw@ust.hk
c e-mail: zwangdq@connect.ust.hk
d e-mail: mzhuan@connect.ust.hk (corresponding author)

## 1 Introduction

The warp drive solution [1] provides a possible way for timelike observers to travel superluminally within the framework of classical General Relativity (GR). In this scenario, the warp bubble, a region of the spacetime deviating from the flat metric, is driven by the local expansion or contraction in its neighboring spacetime. By a proper parameter setting, the propagating speed of the warp bubble can be arbitrarily large, while the internal spacetime of the warp bubble remains quasi-Minkovskian. Hence, the spacecraft inside the warp bubble can travel between two distant points in an arbitrarily short period and remain inside its local lightcones. See [2–5] for recent reviews.

There are several theoretical challenges on the warp drive geometry currently. For example: (i) To generate a warp drive spacetime, exotic matters violating the Weak Energy Condition (WEC) and Null Energy Condition (NEC) [1,6] are required. Superluminal travel and NEC violation are closely related [7,8]. It is realized that the NEC violation is a generic feature of warp drive spacetimes [6]. (ii) The Alcubierre drive requires an unphysically large amount of negative energy, so there appears no practical way to create an Alcubierre drive even if we have control of NEC-violating matter [9,10]. (iii) In the superluminal region, the warp bubble suffers from a "horizon problem": there are event horizons that prevent any observers inside the spaceship to interact with the bubble [11]. (iv) A superluminal warp drive metric leads to quantum instabilities [12–15]. (v) To move the warp drive, the flux of the NEC-violating matter is superluminal. It thus becomes a chicken-and-egg problem of how to move the NEC-violating matter in a superluminal way in order to start the warp drive. One of the main focuses of this field is to solve the above theoretical puzzles, see for example [16–27].

In this paper, we investigate the possibility of superluminal travel through warp drive scenario in another prospective. The above works are concerned with the feasibility of a warp drive metric, while we focus on the potential problems arise in the process superluminal travel. That is, even if we could overcome all puzzles listed above and construct a warp bubble, the additional challenges preventing us from superluminal travel is still severe. We consider two situations that occurs during the interstellar travel. Firstly, a real spacecraft with non-zero volume will feel a "distortional" force due to the nontrivial curvature gradient. Secondly, as indicated by [12], the warp drive metric will accumulate particles in the front part of the bubble, so interstellar matters will be "attracted" and collide with the spacecraft. We study the case for two typical interstellar matter, dusts and photons, and work out the condition for them to collide with the spaceship, the energy shift when the collision happens, and the collision frequency. Then we combine all these results and get the dependence of pressure felt by spaceship on its apparent velocity $v_s$. We find that, although in both cases the pressure depends on $v_s$ linearly when the spaceship is placed at the center, the $v_s$ dependence of photon comes from the energy shift, while that of dust comes from the collision frequency.

The paper is organized as follows. We briefly introduce the warp drive scenario in Sect. 2, and then work out the distortion force exerted on the spaceship in Sect. 3. After that, the geodesic of matter in the warp drive spacetime is studied in Sect. 4, and the total effect on collision between spaceship and interstellar matter is presented in Sect. 5. We finally conclude in Sect. 6.

## 2 The Alcubierre warp drive spacetime

### 2.1 The warp drive metric

The geometry of the Alcubierre warp drive [1] is described as[1]:

$$ds^2 = -c^2 dt^2 + (dx - v_s f(r_s) dt)^2 + dy^2 + dz^2, \quad (2.1)$$

where

$$v_s(t) = \frac{dx_s(t)}{dt}, \quad r_s = \sqrt{(x - x_s(t))^2 + y^2 + z^2}. \quad (2.2)$$

---

[1] Alternative constructions of warp drive spacetime can be found in [28].

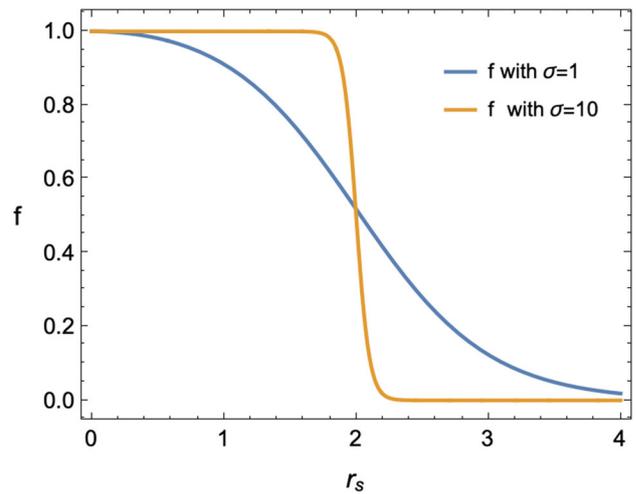

**Fig. 1** $f(r_s)$ for a warp bubble with $R = 2$, $\sigma = 1$ or $\sigma = 10$. Here, $R$ is the radius of the bubble and the parameter $\sigma$ describes the thickness of the bubble. For large $\sigma$, $f$ approaches to a step function

Here, the parameter $r_s$ refers to the distance between any spacetime point $(x, y, z)$ and the center of the bubble. The function $f$ should vanish at a large distance for the spacetime to be asymptotically flat, and becomes a unit at the center of the bubble for the bubble to travel at an apparent velocity $v_s$.[2] A specific expression of $f$ provided by Alcubierre is

$$f(r_s) = \frac{\tanh[\sigma(r_s + R)] - \tanh[\sigma(r_s - R)]}{2\tanh(\sigma R)}, \quad (2.3)$$

and plotted in Fig. 1.

We can understand the warp drive geometry more easily in the limiting case $\sigma \to \infty$. In this limit, $f$ approaches a step function with $f = 1$ for $r_s < R$ and $f = 0$ for $r_s > R$. Recall that $R$ is the radius of the warp bubble, we see the spacetime outside the warp bubble remains quasi-Minkowskian, while the "internal" of the bubble propagates along the $x$ axis with an apparent velocity $v_s$. Since there is no fundamental limit on the parameter $v_s$, we can take it as large as possible, so the warp bubble can take a distant trip with arbitrarily short time.

### 2.2 Frame of reference in the warp drive spacetime

The metric (2.1) describes the global geometry of the spacetime. However, to clarify physical meaning of quantities calculated, a local observer is required. For example, to work out the interaction between particles and the spaceship, we need to parallel-transport their 4-velocities to the same spacetime point and take the inner product in the associated tangent space.

---

[2] In this paper, we consider the case of constant speed $dv_s/dt = 0$ for simplicity.

A natural choice of the local observer is the one associated with the spaceship itself. However, in the local frame of the spaceship, there exist event horizons; moreover, parallel transport of any vector crossing the warp bubble would be mathematically complex. Hence, in this paper, we would prefer to work with an observer remoted from the warp bubble, whose existence is ensured by the asymptotical flatness nature of the metric (2.1).

## 3 Distortion force

Since the warp drive spacetime is curved, we expect a distortion force acting on the spaceship with finite volume. If the gradient of the curvature is large enough, the resulting distortion force may have a significant impact on the spacecraft and then interfere with the superluminal travel. Since in reality, the most important parameter we concern in warp drive metric (2.1) is $v_s$, which determines the efficiency of the superluminal travel, we will fix $R$ and $\sigma$ and determine the dependence of distortion force on $v_s$.

For simplicity, we will work in a "comoving" coordinate $(t, \bar{x}, \rho, \theta)$, related to the original coordinate by

$$\bar{x} = x - v_s t, \ \rho = \sqrt{y^2 + z^2}, \ \theta = \arctan \frac{y}{z}, \qquad (3.1)$$

so $\bar{x}$ is the apparent $x$ coordinate of any spacetime point viewed by the center of the bubble. The apparent velocity is $\bar{v} = d\bar{x}/d\lambda$. The metric is

$$\tilde{g}_{\alpha\beta} = \begin{pmatrix} (v_s^2(f-1)^2 - c^2) & v_s(1-f) & 0 & 0 \\ v_s(1-f) & 1 & 0 & 0 \\ 0 & 0 & 1 & 0 \\ 0 & 0 & 0 & \rho^2 \end{pmatrix}, \qquad (3.2)$$

note that $\theta$ is absent in the metric due to the axial symmetry of warp drive spacetime.

For a realistic interstellar travel, the spaceship had better not to receive too much deformation during the trip. More precisely, we wish for any two neighboring points on the spaceship A and B, their "distance" $r_{AB}$ remains roughly invariant. Here we do not require our spaceship to be a rigid body with constant $r_{AB}$, but to ask the deformation $\Delta r_{AB}$ is small compared to $r_{AB}$. In the comoving frame, this means that the velocity of all the points on the spaceship should vanish, i.e. $dv^\mu/d\lambda \approx 0$ and $v^i \sim 0$. However, the geodesic equation

$$\frac{d\tilde{v}^\mu}{d\lambda} + \Gamma^\mu_{\nu\chi} \tilde{v}^\nu \tilde{v}^\chi = 0, \qquad (3.3)$$

tells that, in the comoving coordinate, any point other than the center of the bubble would have a non-trivial velocity due to the non-zero Christoffel symbol. Hence, there must be an external four-force $F^\mu$ to balance the curvature term

in the geodesic equation. Now, with the external force $F^\mu$, the geodesic equation becomes

$$\frac{d\tilde{v}^\mu}{d\lambda} + \Gamma^\mu_{\nu\chi} \tilde{v}^\nu \tilde{v}^\chi + F^\mu = 0. \qquad (3.4)$$

Note that the distortion of the spaceship comes from the curvature term $\Gamma^\mu_{\nu\chi} \tilde{v}^\nu \tilde{v}^\chi$, so in principle, we shall call this term the distortion force. On the other hand, $F^\mu$ is the force resisting the deformation, which comes from the stiffness of the spaceship. Here for simplicity, we shall call $F^\mu$ as distortion force since, in our case $\frac{d\tilde{v}^\mu}{d\lambda} \simeq 0$ and $F^\mu \simeq -\Gamma^\mu_{\nu\chi} \tilde{v}^\nu \tilde{v}^\chi$.

Since $v^i \sim 0$, the only non-trivial component of (3.4) is $\nu = 0$, $\chi = 0$, which gives

$$F^\mu = \Gamma^\mu_{00}, \qquad (3.5)$$

which is the desired distortion force (or, more precisely, the distortion force density). The geodesic equation (3.4) is valid in any coordinate, and $F^\mu$ is covariant. Thus, $F^\mu$ is invariant under a coordinate transformation, and we can express it in a certain coordinate for simplicity. Here, we choose to work in the comoving coordinate and express $F^\mu$ in the component form. Since the system enjoys axial symmetry, $F^\theta = 0$, and the rest non-trivial spatial components are

$$F^{\bar{x}} = \bar{x} b(r_s) \left( -1 + \frac{1}{2} \coth\left[ R\sigma \right] a(r_s) \right)$$
$$\times \left( 1 - \frac{v_s^2}{c^2} \left( -1 + \frac{1}{2} \coth\left[ R\sigma \right] a(r_s) \right)^2 \right), \qquad (3.6)$$

$$F^\rho = \rho b(r_s) \left( -1 + \frac{1}{2} \coth\left[ R\sigma \right] a(r_s) \right), \qquad (3.7)$$

where we define two auxiliary functions

$$a(r_s) = -\tanh\left[ \sigma(-R + r_s) \right] + \tanh\left[ \sigma(R + r_s) \right], \qquad (3.8)$$

$$b(r_s) = -\frac{v_s^2}{2} \coth\left( R\sigma \right)$$
$$\times \left( \frac{\mathrm{sech}[\sigma(R + r_s)]}{r_s} - \frac{\mathrm{sech}[\sigma(-R + r_s)]}{r_s} \right). \qquad (3.9)$$

We plot the distortion force in Fig. 2 for a practical warp bubble to comprehensively illustrate its behavior. We also plot the $x$ and $\rho$ components of the distortion force as a function of positions in Figs. 3 and 4.

We see from Fig. 3 that $F_x$ is relatively small near $\bar{x} = 0$ and $F_x = 0$ at $\bar{x} = 0$. This can be seen from our auxiliary function $a(r_s)$, since $a(0) = 0$ and $a'(0)$ is suppressed by the hyperbolic functions. The result suggests we put the spaceship in the $y - z$ plane, and the distortion force on $x$ direction would be minimized.

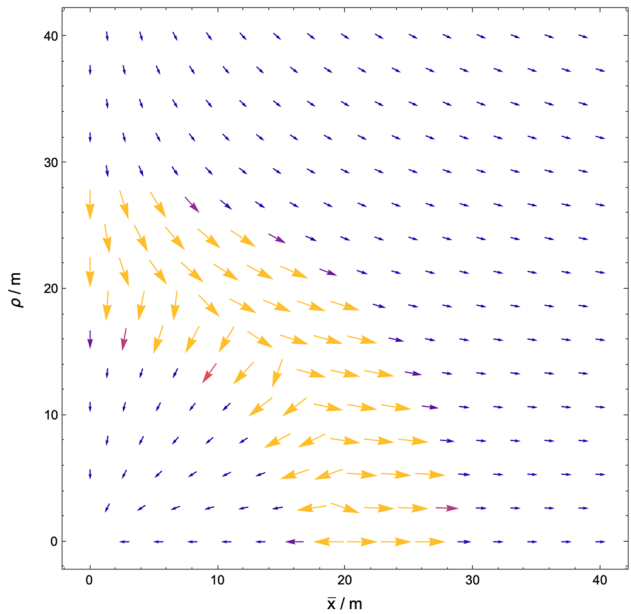

**Fig. 2** Vector plot of the distortion force for a warp bubble with parameters $\sigma = 1$, $R = 20m$, $v_s = 2c$

Figure 4 shows that for a certain warp bubble, $F^\rho$ takes its maximum near $\rho = R$, and its everywhere positive except for two zero points, $\rho \to \infty$ and $\rho = 0$. A detailed proof can be found in Appendix B.

Now we can conclude our result. From Eqs. (3.6) and (3.7) we see $F_x \propto v_s^4$ and $F_\rho \propto v_s^2$. However, $F_x$ can have at most three zero points, one is $\bar{x} = 0$, the other two is expressed by the condition

$$a(r_s) = 2\frac{1 - \frac{c}{v_s}}{\coth[R\sigma]}, \tag{3.10}$$

and the details are in Appendix B. Besides, $F(\rho) = 0$ when $\rho = 0$. Hence, we can always place our spaceship at the center to make distortion force minimized. However, since a real spaceship has a finite volume, it will still feel a non-trivial force which scales as $v_s^2$ and $v_s^4$ in radial and longitudinal direction respectively.

# 4 Motion of particles in the warp drive spacetime

In this section, we work out the motion of particles in the warp drive spacetime, and see whether these particles can collide with the spaceship. We also evaluate the energy shift of these particles. It is generically difficult to evaluate the geodesics with arbitrary initial conditions. However, as we will see later, the collision effect depends linearly on $v_s$ if the spaceship is placed at the center of the bubble, which is small compared to the distortion force when $v_s \gg c$. Hence, if a deviation of the center cannot contribute a $\mathcal{O}(v_s^3)$ factor, then the collision effect would be subdominant compared to the distortion force. Hence, in our preliminary investigation here, we will only work out the case when the spaceship is right at the center of the bubble. A more complete study would be left for further investigation.

## 4.1 Geodesic equations

Given the warp drive metric (2.1), we can work out the corresponding Christoffel symbols, which we present in Appendix A. The geodesic equation is then

$$\frac{d^2 x^\mu}{d\lambda^2} + \Gamma^\mu_{\nu\rho}\frac{dx^\nu}{d\lambda}\frac{dx^\rho}{d\lambda} = 0, \tag{4.1}$$

where $\lambda$ is the affine parameter.

The geodesic equation (4.1) can then be decomposed into four differential equations

$$\ddot{t} + \frac{f^2 f_x v_s^3}{c^2}\dot{t}^2 + \frac{f_x v_s}{c^2}\dot{x}^2 - \frac{2f f_x v_s^2}{c^2}\dot{t}\dot{x} - \frac{f f_y v_s^2}{c^2}\dot{t}\dot{y}$$
$$+ \frac{f_y v_s}{c^2}\dot{x}\dot{y} - \frac{f f_z v_s^2}{c^2}v_s^2\dot{t}\dot{z} + \frac{f_z v_s}{c^2}\dot{x}\dot{z} = 0, \tag{4.2}$$

$$\ddot{x} + \left(\frac{f^3 f_x v_s^4}{c^2} - f f_x v_s^2 - f_t v_s - f\,\partial_t v_s\right)\dot{t}^2$$
$$+ \frac{f f_x v_s^2}{c^2}\dot{x}^2 - \frac{2f^2 f_x v_s^3}{c^2}\dot{t}\dot{x}$$
$$- \left(f^2\frac{v_s^2}{c^2} + 1\right)f f_y v_s\dot{t}\dot{y} - \left(f^2\frac{v_s^2}{c^2} + 1\right)f_z v_s\dot{t}\dot{z}$$
$$+ \frac{f f_y v_s^2}{c^2}\dot{x}\dot{y} + \frac{f f_z v_s^2}{c^2}\dot{x}\dot{z} = 0, \tag{4.3}$$

$$\ddot{y} - f f_y v_s^2\dot{t}^2 + f_y v_s\dot{t}\dot{x} = 0, \tag{4.4}$$

$$\ddot{z} - f f_z v_s^2\dot{t}^2 + f_z v_s\dot{t}\dot{x} = 0, \tag{4.5}$$

where the dot refers to the derivative with respect to the affine parameter $\lambda$. For convenience, we define the 4-velocity of an object as $U^\mu \equiv dx^\mu/d\lambda$. Fortunately, we do not need to go through the full equations. Note that the dust particles should have negligible initial velocity compared to the speed of light in view of a static observer. Hence, in the comoving coordinate, the dust particle should have a conserved angular momentum, perpendicular to the plane determined by the velocity vector $\mathbf{v}_s$ and the line connecting the dust and the center of the bubble. A similar argument also applies to the photon case. The conserved angular momentum ensures that the particles only travel in the plane. Thus, it is sufficient to consider only two coordinates.[3] Hence, we may safely

---

[3] The case is similar to an unbounded trajectory of a point mass around a compact object, e.g. the deflection of light by the Sun. The conservation of angular momentum restricts the movement to be two-dimensional.

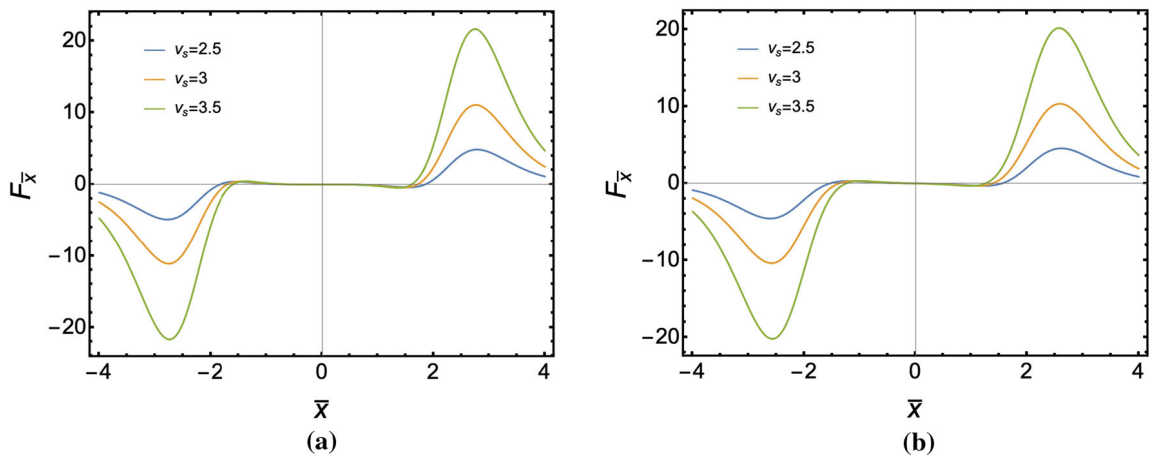

**Fig. 3** The $\bar{x}$ component of the Tidal force with **a** $\rho = 0$, **b** $\rho = 1$ as a function of $\bar{x}$ for different $v_s$. Warp bubble parameters are $R = 2$, $\sigma = 1$ in natural unit

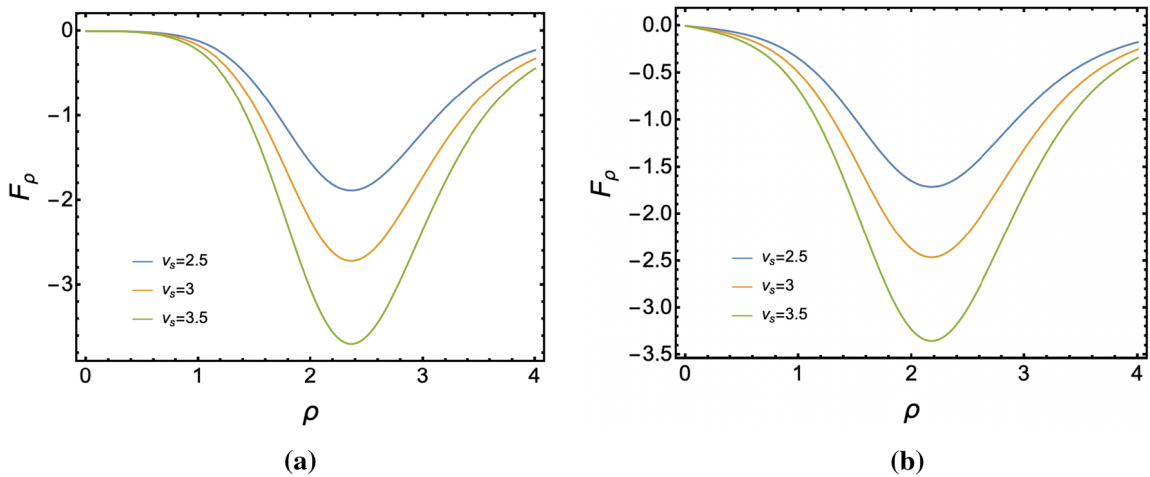

**Fig. 4** The $\rho$ component of the distortion force with **a** $\bar{x} = 0$, **b** $\bar{x} = 1$ as a function of $\rho$ for different $v_s$. Warp bubble parameters are $R = 2$, $\sigma = 1$. All numerical values are in natural units

suppress the $z$ coordinate and take $\dot{z} = 0$ when evaluating equations (4.2) and (4.3).

### 4.2 Dynamics of photons

One main component of interstellar matter is the free photons. In this section we consider the interaction of a single photon with the warp bubble, and determine the energy shift of the photon.

Fortunately, for our concerned case where the spaceship is placed at the center, the trajectory of the photons which could meet with spaceship, as well as the energy shift, i.e., the ratio of photon energy when it hits the spaceship to its energy far from the spaceship, is studied in [29,30]. Here we directly present their results.

Firstly, the apparent angle of the four-velocity of the photon when approaching the spaceship approaches 0. By apparent angle, we mean the angle between the $x$-axis and the four

velocity of the photon, which we illustrate in Fig. 5. This means that

We are also interested in the energy shift of the photon. The result from [29] shows the photon energy shift $\eta_p$ is:

$$\eta_p = \frac{E(r_s = 0)}{E(r_s \gg R)} = 1 + \frac{v_s}{c}, \tag{4.6}$$

where $r_s$ is the apparent distance between the photon and the spaceship viewed from a remote observer. When $v_s \gg c$, the energy shift simplifies to

$$\eta_p \to v_s/c. \tag{4.7}$$

### 4.3 Dynamics of dust matter

Another important matter content in interstellar space is dust. Usually, dust will have a speed much smaller than the speed

**Fig. 5** The original scenario where the particles propagate from infinity to the center of the bubble, shown in **a**, is equivalent to that in **b**, where a particle is emitted from the center of the bubble. The photon cannot go outside the bubble unless the apparent angle $\zeta$ approaches 0

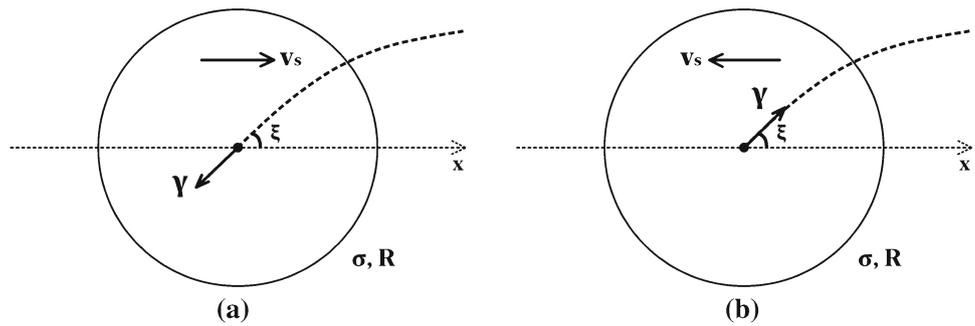

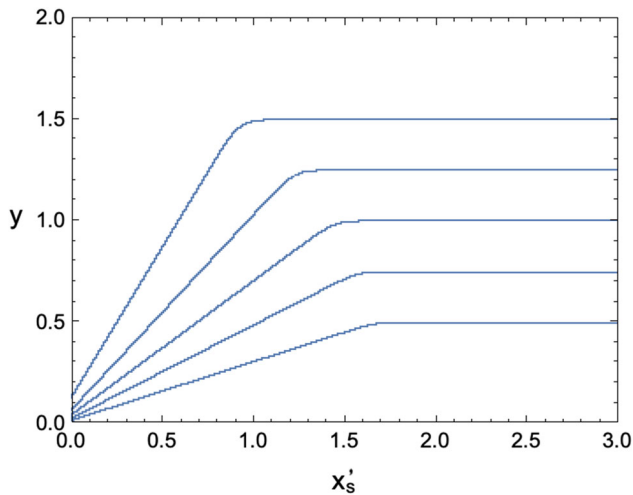

**Fig. 6** The trajectories of dust particles with a large initial $x$ position, and initial $y$ positions at $y = 0.5$, $y = 0.75$, $y = 1$, $y = 1.25$ and $y = 1.5$ respectively. The parameters of the warp bubble read $\sigma = 10$, $R = 2$, $v_s = 100$. All numerical values are in natural units. Here, $x_s' \equiv x - v_s t$ represents the apparent distance of dust and spaceship in $x$ axis, so $x_s' = 0$ means spaceship and dust have the same $x$ coordinate in the view of the remote observer

of light, so for a warp bubble with superluminal speed, we can safely set their speed to zero when they are remote from the bubble. Hence, the boundary condition for the geodesic equation is set to be

$$U^\mu(r \to \infty) = (1, 0, 0, 0), \tag{4.8}$$

and for each dust particle with $y$ coordinate $\lim_{r\to\infty} y = y_0$, we can uniquely determine its geodesics.

We numerically evaluate the geodesic equation with different $y_0$, and the trajectories are illustrated in Fig. 6. We see the dust particles entering the warp bubble from rest will tend to move towards the center of the bubble, but will not exactly reach the center. The dust particles with a larger initial $y$ position will arrive at the x–z plane further away from the spaceship. This is consistent with the numerical result by [30].

One may notice that, in Fig. 6, the point where the trajectory of the dust particles starts to shift is inside the bubble. This comes from the fact that, with respect to the bubble, the particle carries a large initial velocity along x-direction, while it takes some time to accelerate along y-direction.

We can comprehensively understand the result in Fig. 6 in another way. It is found that in a generic warp drive spacetime, the Ricci scalar has the greatest impact on the surrounding spacetime [31,32]. This reaches an agreement with our result. One may find from Fig. 6 that the acceleration of the dust reaches its maximum near the boundary of the warp drive bubble, where the trajectory starts to bend. The maximum value of the gradient of the Ricci scalar is also near the boundary of the bubble. This means that dust acceleration is positively correlated to the gradient of the Ricci scalar.

In order to calculate the energy of the collision, we need the four velocity of the particle at the collision event. The four velocity can be calculated from the apparent velocity of the particle when arriving at the center of the bubble, i.e., the apparent velocity of the particle viewed from the spaceship, which we will call the final velocity for simplicity. To study the properties of the final velocity, we numerically integrated the geodesics equations (4.2) and (4.3) for a dust particle. This particle starts at rest far away from the warp bubble. Its initial velocity is set by Eq. (4.8). As discussed above, for a spaceship located at the center of the bubble, dusts with initial coordinate $y_0 \neq 0$ is hard to hit the spaceship. Thus, it suffices to consider the dusts with $y_0 \simeq 0$ only. We depict the result in Fig. 7. We see that under the high-speed condition, the x-component of the final velocity of these dust particles approaches to $v_x = v_s - c$. In the frame of a remote observer, the four velocity of the particle can be formally written as

$$v^\alpha = (\gamma, \gamma v_x, 0, 0). \tag{4.9}$$

The Eq. (4.9), along with the normalization of $v^\alpha$

$$v^\alpha v_\alpha = -1, \tag{4.10}$$

gives the expression

$$v^\alpha = \left( \frac{c}{\sqrt{v_s^2 - 2v_s c}} i, \frac{c(v_s - c)}{\sqrt{v_s^2 - 2v_s c}} i, 0, 0 \right). \tag{4.11}$$

Note that the four velocity is imaginary, this is because the particles, when inside the bubble, travels along a spacelike

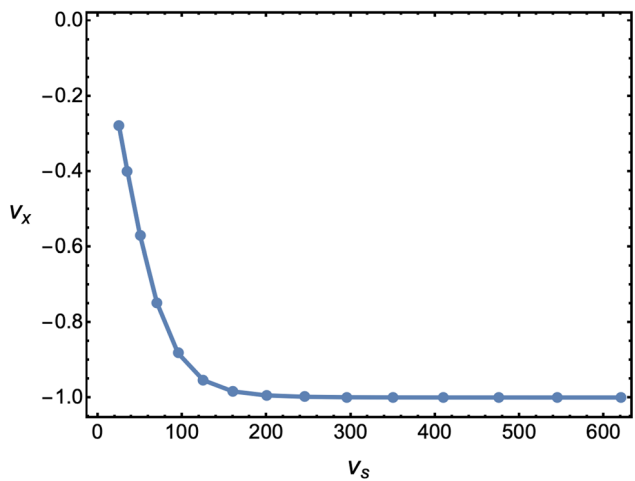

**Fig. 7** The final apparent velocity minus the speed of the bubble as a function of the speed of the bubble, measured in a remote observer. The dust particle starts at rest at x = 2.5, y = 0, and the parameters of the bubble are R = 2, $v_s$ = 20, and $\sigma$ = 10. All numerical values are in natural units. The center of the bubble is placed at x = 0 at start

geodesics viewed by a remote observer. Similarly, since the apparent velocity of the spaceship is $v_s$, we can work out its corresponding four-velocity

$$u^\alpha = \left( \frac{c}{\sqrt{v_s^2 - c^2}} i, \frac{v_s c}{\sqrt{v_s^2 - c^2}} i, 0, 0 \right). \qquad (4.12)$$

The energy of the dust particle per unit mass seen by an observer on the spaceship is thus:

$$E_d = -u^\alpha v_\alpha = \frac{v_s^2 c^2 - c^4 - v_s c^3}{\sqrt{v_s^2 - c^2} \sqrt{(v_s - c)^2 - c^2}}, \quad \lim_{v_s \to \infty}$$
$$E_d = c^2. \qquad (4.13)$$

Note that, the energy is a scalar coming from the inner product of two vector, so its value is independent of the reference frame we choose, so we can evaluate it in a remote observer for simplicity.

We can conclude from the above result that, under high speed condition $v_s \gg c$, a dust particle entering the warp bubble will interact with the spaceship with energy $E_d \propto v_s^0$.

Finally, for completeness, we also plot the final velocity of a dust particle entering the bubble with initial coordinate $y \neq 0$ in Fig. 8. Similar to the above case, the final velocity approaches a certain value for large $v_s$. We can read from the plots that, as the initial y position increases, the final velocity along the x-axis decreases. Thus, the head-on collision between the dust particle and the spaceship is weakened. Besides, the combination of Fig. 8a, b tells that, the total apparent velocity of the dust particle approaches the speed of light for large $v_s$. Hence, the condition $v_x = v_s - c$ for the $y_0 = 0$ case can be regarded as a specific case with $v_y = 0$.

## 5 The pressure caused by photons and dust particles in the interstellar space

Now that we get the trajectory of photons and dust particles as well as their energy shift at collision event, the rest thing we shall do is to work out the frequency of the collision. After that, we can get the dependence of collisions on $v_s$.

### 5.1 Photon collision

Since the photons colliding with spaceship at the center will all have a vanishing apparent angle $\zeta$, we may take the following simplification. We take $\sigma$ to be large enough, such that the spacetime when $r_s < R$ and $r_s > R$ are almost flat. Hence, the photons colliding with the spaceship must enter the bubble through the neighborhood of the spatial point $(v_s t, 0, 0)$, the area of the neighborhood denoted as $\Delta A$. Moreover, since light rays emitted at $\Delta A$ with any incidence angle will all have a vanishing apparent angle inside the bubble, the photons colliding with the spaceship at a certain time must originate from a spherical shell. We illustrate the above point in Fig. 9.

Since the position and velocity of interstellar photons are randomly distributed, we see that the collision frequency should be proportional to the differential volume of the photon layer. In this case, the total momentum $dp_p$ of photons travelling towards the area $\Delta A$ in an infinitesimal time $dt$ comes from a certain solid angle $d\Omega$ is:

$$dp_p' = p_{av}(\rho_p \, dV) \left( \frac{\Delta A \cos\theta}{4\pi R_p^2} \right)$$
$$= p_{av} c \rho_p \left( \frac{\Delta A \cos\theta}{4\pi} \right) d\Omega \, dt, \qquad (5.1)$$

where $p_{av}$ is the average momentum value of the interstellar photons. For each photon through the bubble, the blue-shift causes its energy to increase by $\eta_p \sim v_s/c$, so the total momentum of photons at the bridge within time $dt$ is:

$$dp_p = \eta_p \int_\Omega dp_p' = \frac{E_{av}}{4} \rho_p \Delta A \left( \frac{v_s}{c} \right) dt. \qquad (5.2)$$

Here $E_{av}$ is the average energy of interstellar photons $E_{av}$. The interaction of the high-energy photons with the spacecraft will lead to the momentum transfer, causing the light pressure $P_p$ on the foredeck.

$$P_p = (1 + n) \left( \frac{dp_p}{dt} \right) \frac{1}{\Delta A} = \frac{(1 + n)}{4} \rho_p E_{av} \left( \frac{v_s}{c} \right) \propto v_s, \qquad (5.3)$$

where n is the reflection coefficient of the spacecraft foredeck. We can thus conclude from Eq. (5.3) that the pressure

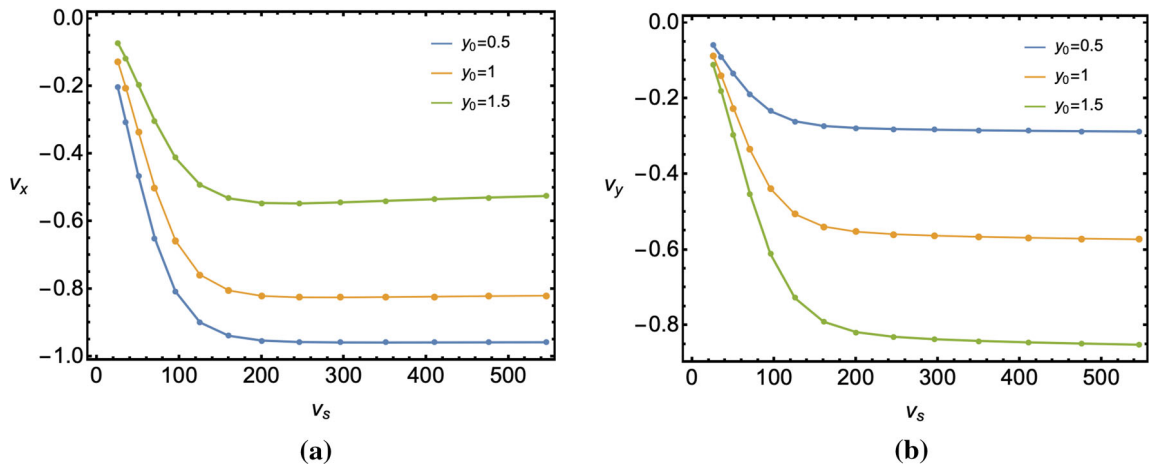

**(a)**                                                                                  **(b)**

**Fig. 8** The apparent velocity as a function of $v_s$, measured in a remote observer. The parameter of the bubble read $R = 2$, $v_s = 20c$, and $\sigma = 10$. The center of the bubble is placed at $x = 0$ at start. Figure 7a shows the x component of the velocity of the dust particle at the center minus the speed of the bubble as a function of $v_s$, with initial position of the particle to be $x = 2.5$ and $y_0 = 0.5, 1.0, 1.5$ respectively. Figure 7b shows the y component of the velocity of the dust particle at the center as a function of $v_s$, with initial position of the particle to be $x = 2.5$ and $y = 0.5, 1.0, 1.5$ respectively. All the velocities here are the apparent velocities measured by a remote observer. All numerical values are in natural units

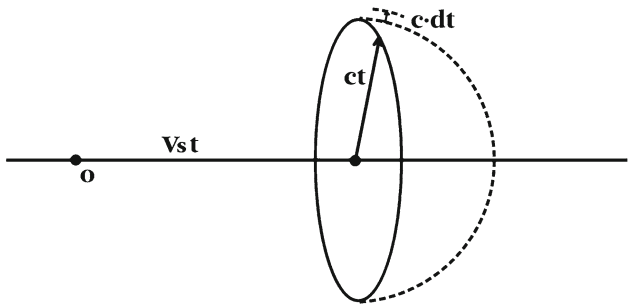

**Fig. 9** The photon layer. The photons which will arrive at the center of the bubble after time t are laying on a shell with radius $R = ct$ and thickness $cdt$. The distance between the center of the bubble and the center of the shell is $v_s t$

from photon collision is proportional to the spacecraft apparent velocity $v_s$.

### 5.2 Dust particle collision

The velocity of interplanetary dust particles, which is typically around $10^{-4}c$, is extremely small compared to the velocity of our warp bubble. The dust particles that hit the warp bubble will therefore be mostly those staying on the cylinder that our warp bubble swaps through. At the same time, since the spaceship is much smaller than the warp bubble, we consider our spaceship as a point sitting at the center of the bubble. In this case, the total momentum $dp_d$ of dust particles entering the warp bubble in an infinitesimal time $dt$

is:

$$dp_d = p'_{av}\rho_d \pi \Delta A v_s dt \tag{5.4}$$

where $p'_{av}$ is the average momentum value of dust particles. The average pressure caused by dust particles at the bridge is thus:

$$P_d = (1+n)E_d\left(\frac{dp_d}{dt}\right)\frac{1}{\Delta A} \propto v_s \tag{5.5}$$

where $n$ is the reflection coefficient of the spacecraft front deck. We see the collision with photons and dust particles will both contribute a pressure proportional to $v_s$.

## 6 Conclusion and outlook

We considered two possible obstacles on the warp drive spacecraft through interstellar travel and calculated their dependence on the speed of the warp bubble $v_s$. Our results shows that when $v_s \gg c$, the distortion force scales as $v_s^2$ and $v_s^4$ in radial and longitudinal direction respectively, and would be minimized when we place the spaceship in the center of the warp bubble. The pressure on the front deck of the spacecraft caused by the interstellar photons and dust particles collision is proportional to $v_s$, and hence subdominant compared to the distortional force.

Our work can be extended in various directions. Firstly, we may consider a warp bubble with varying speed $v_s$, and see whether non-trivial accelerations may lead to new phenomenons. Secondly, investigations on the collision effect when spaceship is not placed at the center may be interesting. Thirdly, it is interesting to study whether dark matter or even dark energy can interact with the warp bubble as well as the spaceship. Finally, a stability check, on both classical and quantum levels, is worth studying.

**Acknowledgements** We thank Qianhang Ding, Chon Man Sou, Xi Tong, Shengfeng Yan, Jinbo Yang and Siyi Zhou for helpful discussions. This work was supported in part by the National Key R&D Program of China (2021YFC2203100), and the NSFC Excellent Young Scientist Scheme (Hong Kong and Macau) Grant no. 12022516.

Funded by SCOAP³. SCOAP³ supports the goals of the International Year of Basic Sciences for Sustainable Development.

## Appendix A: Christoffel symbols

In this appendix, we provide all the non-zero Christoffel symbols under the warp drive metric $g_{\alpha\beta}$:

$$\Gamma^t_{tt} = \frac{1}{c^2}(f_x f^2)v_s^3, \quad \Gamma^t_{xt} = -\frac{1}{c^2}(f_x f)v_s^2,$$

$$\Gamma^t_{yt} = -\frac{1}{2c^2}(f_y f)v_s^2, \quad \Gamma^t_{xx} = \frac{1}{c^2}f_x v_s, \tag{A.1}$$

$$\Gamma^t_{xz} = \frac{1}{2c^2}f_z v_s, \quad \Gamma^t_{xy} = \frac{1}{2c^2}f_y v_s,$$

$$\Gamma^t_{tz} = -\frac{1}{2c^2}(f_z f)v_s^2, \tag{A.2}$$

$$\Gamma^x_{tx} = -\frac{1}{c^2}(f_x f^2)v_s^3, \quad \Gamma^x_{xx} = \frac{1}{c^2}(f_x f)v_s^2,$$

$$\Gamma^x_{xz} = \frac{1}{2c^2}(f_z f)v_s^2, \Gamma^x_{xy} = \frac{1}{2c^2}(f_y f)v_s^2, \tag{A.3}$$

$$\Gamma^x_{ty} = -\frac{f_y v_s}{2} - \frac{(f_y f^2 v_s^3)}{2c^2}, \quad \Gamma^x_{tz} = -\frac{f_z v_s}{2} - \frac{(f_z f^2 v_s^3)}{2c^2}, \tag{A.4}$$

$$\Gamma^x_{tt} = \frac{(f_x f^3 v_s^4)}{c^2} - v_s(f_t + v_s f_x f) - f \partial_t v_s, \tag{A.5}$$

$$\Gamma^y_{tt} = -f_y f v^2, \quad \Gamma^y_{tx} = \frac{f_y v}{2},$$

$$\Gamma^z_{tt} = -f_z f v^2, \quad \Gamma^z_{tx} = \frac{f_z v}{2}. \tag{A.6}$$

## Appendix B: Stable surface under the distortion force

In this appendix, we will show that we cannot find any surface where $F^\rho = 0$ except for the trajectory where $\rho = 0$, but a surface with $F^{\bar{x}} = 0$ is possible. We define a new function $n(\sigma, r_s)$ as:

$$n(\sigma, r_s) = \tanh[\sigma R - \sigma r_s] + \tanh[\sigma R + \sigma r_s]$$
$$- 2\tanh[\sigma R]. \tag{B.1}$$

Now that $r_s < R$ for our spaceship, and since $\tanh x$ is a concave function for $x > 0$, we immediately know that $n(\sigma, r_s) < 0$, which means

$$-1 + \frac{1}{2}\coth[R\sigma]a(r_s) < 0. \tag{B.2}$$

Thus, the only place where $F^\rho = 0$ is along the $\bar{x}$ axis. However, we can make $\bar{x}$ component of the tidal force zero, $F^{\bar{x}} = 0$, by imposing:

$$c^2 - v_s^2\left(-1 + \frac{1}{2}\coth[R\sigma]a(r_s)\right)^2 = 0. \tag{B.3}$$

Together with Eq. (B.2), we obtain the requirement for a sphere with zero $F^{\bar{x}}$:

$$a(r_s) = 2\frac{1 - \frac{c}{v_s}}{\coth[R\sigma]}. \tag{B.4}$$

## References

1. M. Alcubierre, The Warp drive: hyperfast travel within general relativity. Class. Quantum Gravity **11**, L73–L77 (1994). https://doi.org/10.1088/0264-9381/11/5/001. arXiv:gr-qc/0009013
2. Invited chapter to appear in an edited collection 'Classical and Quantum Gravity Research Progress', Nova Science Publishers
3. M. Alcubierre, F.S.N. Lobo, Warp drive basics. Fundam. Theor. Phys. **189**, 257–279 (2017). https://doi.org/10.1007/978-3-319-55182-1_11
4. M. Alcubierre, *Wormholes, Warp Drives and Energy Conditions*, vol. 189 (Springer, Berlin, 2017). https://doi.org/10.1007/978-3-319-55182-1. arXiv:2103.05610 [gr-qc]
5. B. Shoshany, Lectures on faster-than-light travel and time travel. SciPost Phys. Lect. Notes **10**, 1 (2019). https://doi.org/10.21468/SciPostPhysLectNotes.10. arXiv:1907.04178 [gr-qc]
6. F.S.N. Lobo, M. Visser, Fundamental limitations on "warp drive" spacetimes. Class. Quantum Gravity **21**, 5871–5892 (2004). https://doi.org/10.1088/0264-9381/21/24/011. arXiv:gr-qc/0406083
7. K.D. Olum, Superluminal travel requires negative energies. Phys. Rev. Lett. **81**, 3567–3570 (1998). https://doi.org/10.1103/PhysRevLett.81.3567. arXiv:gr-qc/9805003

8. M. Visser, B. Bassett, S. Liberati, Superluminal censorship. Nucl. Phys. B Proc. Suppl. **88**, 267–270 (2000). https://doi.org/10.1016/S0920-5632(00)00782-9. arXiv:gr-qc/9810026

9. M.J. Pfenning, L.H. Ford, The Unphysical nature of 'warp drive'. Class. Quantum Gravity **14**, 1743–1751 (1997). https://doi.org/10.1088/0264-9381/14/7/011. arXiv:gr-qc/9702026

10. L.H. Ford, T.A. Roman, Restrictions on negative energy density in flat space-time. Phys. Rev. D **55**, 2082–2089 (1997). https://doi.org/10.1103/PhysRevD.55.2082. arXiv:gr-qc/9607003

11. S.V. Krasnikov, Hyperfast travel in general relativity. Phys. Rev. D **57**, 4760–4766 (1998). https://doi.org/10.1103/PhysRevD.57.4760. arXiv:gr-qc/9511068

12. S. Finazzi, S. Liberati, C. Barcelo, Semiclassical instability of dynamical warp drives. Phys. Rev. D **79**, 124017 (2009). https://doi.org/10.1103/PhysRevD.79.124017. arXiv:0904.0141 [gr-qc]

13. A. Coutant, S. Finazzi, S. Liberati, R. Parentani, Impossibility of superluminal travel in Lorentz violating theories. Phys. Rev. D **85**, 064020 (2012). https://doi.org/10.1103/PhysRevD.85.064020. arXiv:1111.4356 [gr-qc]

14. B. McMonigal, G.F. Lewis, P. O'Byrne, The Alcubierre warp drive: on the matter of matter. Phys. Rev. D **85**, 064024 (2012). https://doi.org/10.1103/PhysRevD.85.064024. arXiv:1202.5708 [gr-qc]

15. S. Liberati, Do not mess with time: probing faster than light travel and chronology protection with superluminal warp drives, in *14th Marcel Grossmann Meeting on Recent Developments in Theoretical and Experimental General Relativity, Astrophysics, and Relativistic Field Theories*, vol. 2 (2017), pp. 1407–1414. https://doi.org/10.1142/9789813226609_0120. arXiv:1601.00785 [gr-qc]

16. C. Van Den Broeck, A "Warp drive" with reasonable total energy requirements. Class. Quantum Gravity **16**, 3973–3979 (1999). https://doi.org/10.1088/0264-9381/16/12/314. arXiv:gr-qc/9905084

17. C. Barcelo, S. Finazzi, S. Liberati, On the impossibility of superluminal travel: the warp drive lesson. arXiv:1001.4960 [gr-qc]

18. E.W. Lentz, Breaking the warp barrier: hyper-fast solitons in Einstein–Maxwell-plasma theory. Class. Quantum Gravity **38**(7), 075015 (2021). https://doi.org/10.1088/1361-6382/abe692. arXiv:2006.07125 [gr-qc]

19. G.U. Varieschi, Z. Burstein, Conformal gravity and the Alcubierre warp drive metric. ISRN Astron. Astrophys. **2013**, 482734 (2013). https://doi.org/10.1155/2013/482734. arXiv:1208.3706 [gr-qc]

20. A. DeBenedictis, S. Ilijic, Energy condition respecting warp drives: the role of spin in Einstein–Cartan theory. Class. Quantum Gravity **35**(21), 215001 (2018). https://doi.org/10.1088/1361-6382/aae326. arXiv:1807.09745 [gr-qc]

21. O.L. Santos-Pereira, E.M.C. Abreu, M.B. Ribeiro, Fluid dynamics in the warp drive spacetime geometry. Eur. Phys. J. C **81**(2), 133 (2021). https://doi.org/10.1140/epjc/s10052-021-08921-3. arXiv:2101.11467 [gr-qc]

22. O.L. Santos-Pereira, E.M.C. Abreu, M.B. Ribeiro, Charged dust solutions for the warp drive spacetime. Gen. Relativ. Gravit. **53**(2), 23 (2021). https://doi.org/10.1007/s10714-021-02799-y. arXiv:2102.05119 [gr-qc]

23. A. Bobrick, G. Martire, Introducing physical warp drives. Class. Quantum Gravity **38**(10), 105009 (2021). https://doi.org/10.1088/1361-6382/abdf6e. arXiv:2102.06824 [gr-qc]

24. S.D.B. Fell, L. Heisenberg, Positive energy warp drive from hidden geometric structures. Class. Quantum Gravity **38**(15), 155020 (2021). https://doi.org/10.1088/1361-6382/ac0e47. arXiv:2104.06488 [gr-qc]

25. J. Santiago, S. Schuster, M. Visser, Generic warp drives violate the null energy condition. Phys. Rev. D **105**(6), 064038 (2022)

26. M. Visser, J. Santiago, S. Schuster, Tractor beams, pressor beams, and stressor beams within the context of general relativity, in *Contribution to the Sixteenth Marcel Grossman Conference (MG16, Rome, July, 2021)*. arXiv:2110.14926 [gr-qc]

27. E.W. Lentz, Hyper-fast positive energy warp drives, in *Contribution to the Sixteenth Marcel Grossman Conference (MG16, Rome, Italy, July 2021)*. arXiv:2201.00652 [gr-qc]

28. J. Natário, Warp drive with zero expansion. Class. Quantum Gravity **19**, 1157–1166 (2002). https://doi.org/10.1088/0264-9381/19/6/308. arXiv:gr-qc/0110086

29. C. Clark, W.A. Hiscock, S.L. Larson, Null geodesics in the Alcubierre warp drive space-time: the view from the bridge. Class. Quantum Gravity **16**, 3965–3972 (1999). https://doi.org/10.1088/0264-9381/16/12/313. arXiv:gr-qc/9907019

30. T. Muller, D. Weiskopf, Detailed study of null and time-like geodesics in the Alcubierre Warp spacetime. Gen. Relativ. Gravit. **44**, 509–533 (2012). https://doi.org/10.1007/s10714-011-1289-0. arXiv:1107.5650 [gr-qc]

31. B. Mattingly et al., Curvature invariants for the accelerating Natário warp drive. Particles **3**(3), 642–659 (2020). https://doi.org/10.3390/particles3030042. arXiv:2008.03366 [gr-qc]

32. B. Mattingly et al., Curvature invariants for the Alcubierre and Natário warp drives. Universe **7**(2), 21 (2021). https://doi.org/10.3390/universe7020021. arXiv:2010.13693 [gr-qc]